\documentclass{iau} 
\usepackage{graphicx}

\title[The IMF of young open clusters] %% give here short title %%
{The Initial Mass Function of Young Open Clusters in the Galaxy: A Preliminary Result}

\author[Lim et al.]   %% give here short author list %%
{Beomdu Lim$^1$,
 Hwankyung Sung$^2$, Hyeonoh Hur$^2$, \\ \and Byeong-Gon Park$^{1,3}$}

\affiliation{$^1$Korea Astronomy and Space Science Institute, 776 Daedeokdae-ro, Yuseong-gu, Daejeon 305-348, Korea \\ email: {\tt bdlim1210@kasi.re.kr} \\[\affilskip]
$^2$Dept. of Astronomy \& Space Science, Sejong University, 209 Neungdong-ro, Gwangjin-gu, Seoul 143-747, Korea \\
$^3$Korea University of Science and Technology, Gajeong-ro, Yuseong-gu, Daejeon 305-333, Korea \\}

\pubyear{2015}
\volume{316}  %% insert here IAU Symposium No.
\jname{Formation, evolution, and survival of massive star clusters}
\editors{C. Charbonnel \& A. Nota, eds.}
\begin{document}

\maketitle

\begin{abstract}
The initial mass function (IMF) is an essential tool with which to study star formation processes. We have 
initiated the photometric survey of young open clusters in the Galaxy, from which the stellar IMFs are obtained 
in a homogeneous way. A total of 16 famous young open clusters have preferentially been studied up to now. 
These clusters have a wide range of surface densities ($\log \sigma = -1$ to 3 [stars pc$^{-2}$] for stars with mass 
larger than 5$M_{\odot}$) and cluster masses ($M_{\mathrm{cl}} = 165$ to $50,000 M_{\odot}$), and also 
are distributed in five different spiral arms in the Galaxy. It is possible to test the dependence of star formation 
processes on the global properties of individual clusters or environmental conditions. We present a preliminary 
result on the variation of the IMF in this paper.

\end{abstract}

\firstsection % if your document starts with a section,
\section{Introduction}
It is well recognized that the evolution of stars mainly depends on their initial mass, partly on 
the chemical composition. However what determines the stellar mass is poorly understood. The origin 
of the stellar mass spectrum is thought to be closely related to star formation processes. And therefore 
the initial mass function (IMF), the mass distribution of stars from a single star formation event, is a 
crucial diagnostic tool with which to understand star formation processes.  

Although the IMF seems to be universal according to a number of previous studies (\cite[Bastian, Covey, \& Meyer 2010]{BCM10}), 
many evidences for the variation of the IMF have been reported both from observational and theoretical studies (\cite[Lu 
et al. 2013]{LDG13}; \cite[Marks et al. 2012]{MKD12}; etc). One of main reasons for this debate may be caused by 
the use of inhomogeneous data. Thus, we focus on the results from a homogeneous imaging survey of young open 
clusters in the Galaxy (``Sejong Open cluster Survey'' -- \cite[Sung et al. 2013]{SBL13}).

\section{Preliminary Results and Conclusions}
\begin{figure}[t]
% \vspace*{-2.0 cm}
\begin{center}
 \includegraphics[width=3.5in]{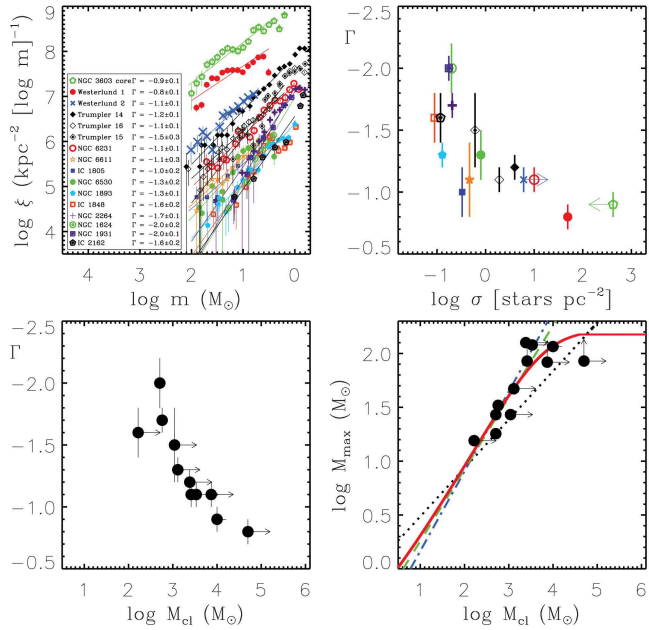} 
% \vspace*{-1.0 cm}
 \caption{The initial mass functions (IMF) of 16 young open clusters in the Galaxy (upper left), variations of the IMF 
slope $\Gamma$ with respect to the surface density of the clusters (upper right) and the cluster mass (lower left), and 
a dependency of maximum stellar masses on the mass of host clusters (lower right). In the lower right-hand panel, the 
dot-dashed, dotted, dashed, and solid lines show $M_{\mathrm{max}}$-$M_{\mathrm{cl}}$ relations from analytic 
approaches of \cite[Elmegreen (2000)]{E00}, \cite[Larson (2003)]{L03}, \cite[Bonnell, Bate, \& Vine (2003)]{BBV03}, 
and \cite[Weidner \& Kroupa (2004)]{WK04}, respectively. }
   \label{fig1}
\end{center}
\end{figure}

The young open clusters are useful objects in the study of the IMF because they are not only a coeval and co-spatial 
population but also an ecological indicator in the Galaxy. We investigated the IMF of 16 famous young open clusters by 
analyzing their Hertzsprung-Russell diagrams. The upper left-hand panel of Fig.\,\ref{fig1} shows their IMF and 
its slope $\Gamma$ (the data from \cite[Lim 2014]{L14}, references therein, and ongoing work).  

Variations of the slope $\Gamma$ with respect to the surface density and total mass of the clusters are shown in 
the upper right and lower left-hand panels, respectively. The slope of the IMF in the high-mass 
regime ($> 1 M_{\odot}$) appears to be shallow for massive dense clusters. The maximum stellar mass is 
also larger for more massive clusters (lower right-hand panel). This $M_{\mathrm{max}}$-$M_{\mathrm{cl}}$ relation is 
consistent with results from analytic approaches (\cite[Elmegreen 2000]{E00}; \cite[Larson 2003]{L03}; 
\cite[Bonnell, Bate, \& Vine 2003]{BBV03}; \cite[Weidner \& Kroupa 2004]{WK04}). It may imply the deterministic 
origin of the stellar mass.

The Galactic variations of the IMF and the cluster properties were also investigated to simply test the environmental 
influences on the star formation processes. The massive dense clusters tend to be formed in the inner Galaxy, while 
the formation of less massive sparse clusters seem to be dominant in the outer Galaxy. We tentatively suggest that the star formation is likely 
to be controlled by local environmental conditions rather than a universal process.

\end{document}